\documentclass[twocolumn,english,aps,prl,superscriptaddress,longbibliography]{revtex4-1}
\usepackage[T1]{fontenc}
\usepackage[latin9]{inputenc}
\usepackage{amsmath}
\usepackage{amssymb}
\usepackage{graphicx}

\makeatletter

\usepackage{epsfig}
\usepackage{dcolumn}
\usepackage{bm}
\usepackage{bbm}
\usepackage{amsfonts}
\usepackage{latexsym}
\usepackage{color}
\usepackage[normalem]{ulem}

\makeatother

\usepackage{babel}
\begin{document}

\title{Hidden Chern number in one-dimensional non-Hermitian chiral-symmetric systems}

\author{Wojciech Brzezicki}

\affiliation{International Research Centre MagTop, Institute of Physics, Polish
Academy of Sciences,~\\
 Aleja Lotnikow 32/46, PL-02668 Warsaw, Poland}

\author{Timo Hyart}

\affiliation{International Research Centre MagTop, Institute of Physics, Polish
Academy of Sciences,~\\
 Aleja Lotnikow 32/46, PL-02668 Warsaw, Poland}
\begin{abstract}
We consider a class of one-dimensional non-Hermitian models with a
special type of a chiral symmetry which is related to pseudo-Hermiticity.  We show that the topology of a Hamiltonian belonging to this symmetry class is determined by a hidden Chern number described by an effective 2D Hermitian Hamiltonian $H^{\rm eff} (k, \eta)$, where $\eta$ is the imaginary part of the energy. This Chern number manifests itself as topologically protected in-gap
end states at zero real part of the energy. We show that the bulk-boundary
correspondence coming from the hidden Chern number is robust and immune
to non-Hermitian skin effect. We introduce a minimal  model  Hamiltonian supporting topologically nontrivial phases in this symmetry class, derive its topological phase diagram and calculate the end states originating from the hidden Chern number.
\end{abstract}
\maketitle

 Open quantum systems with loss (dissipation) and gain (coherent amplification)  are described by non-Hermitian (NH) Hamiltonians \cite{Nature-Phys-Exp18, Konotop16RMP} and have unexpected properties which often depend on the symmetries of the system \cite{Nature-Phys-Exp18, Konotop16RMP, Ben07, Hatano-Nelson, Efetov97, Bernard-LeClair, Bernard-LeClair2, Magnea08, Sato12-review, Deng12, Pikulin14}. 
 Adding a non-Hermitian component to a Hamiltonian does not only broaden the resonances and allow the eigenstates to decay, but the eigenmodes can merge with each other at exceptional points, which are topological defects  where not only the eigenvalues are degenerate but also the eigenvectors are parallel to each other \cite{Nature-Phys-Exp18, Miri19, Berry04, Heiss12, Lui19}. The flexibility to engineer gain and loss in a controllable manner for example in optics, 
(opto)mechanics, plasmonics, superconducting quantum circuits, dissipative Bose-Einstein condensates, exciton-polariton condensates and cold atom systems 
 \cite{Nature-Phys-Exp18, Konotop16RMP, Miri19, Muller, Aspelmeyer14, Cao15, Gao15, Zeu15, Schomerus15, Xu16, Wei17, Xiao17, Fitzpatrick17, Zhou18, Wang18, Cer18, Silveri18, Silveri19a, Silveri19b,Yosh19} have naturally raised the interest to study the symmetry and topology in non-Hermitian physics systematically \cite{Gong18, Lieu18, Zhou19, Kaw19, Gha19} with potential applications for example in the design of topologically protected laser modes \cite{Bah17, Jean17, Par18, Zhao18, Har18, Ban18}.

Because in NH systems the energies are complex the Altland-Zirnbauer symmetry classes support  new types of winding number and $\mathbb{Z}_{2}$ invariants determined by the complex spectra \cite{Gong18} -- leading to NH topological phases with no Hermitian counterparts \cite{Gong18, Zhou19, Kaw19}.
The classification is further enriched because for a NH Hamiltonian the transpose and complex conjugation are not equivalent so that the 10 Altland-Zirnbauer symmetry classes need to be extended \cite{Lieu18} to 38 non-Hermitian (nonspatial) symmetry classes   \cite{Zhou19, Kaw19}. Furthermore, Hermitian Hamiltonians are gapped if the energy bands do not cross the Fermi energy, but non-Hermitian systems  feature two different types of complex-energy gaps, so-called point (line) gaps where bands do not cross a point (line) in the complex-energy plane, giving rise to further ramification of the topological classification \cite{Kaw19}. Various models and realizations of the different NH topological phases have been proposed 
\cite{Nelson98, Rudner-Levitov, Esa11, Lia13, Schomerus13, Rud16, Lee16, Ley17,  Jin17, Lieu18B, Shen18, Xu17, Tak18, Wang19, Hui19}. 
For Hermitian Hamiltonians the bulk-boundary correspondence guarantees that the topological invariants for periodic boundary conditions predict the presence of boundary states for open boundary conditions, but this is typically not the case for NH systems \cite{Xio18,Zirn19}, where the experimental consequences of topological invariants are less clear because the bulk-boundary correspondence can typically be established 
either on  the level of singular value spectra \cite{Herv19} or on the level of biorthogonal density \cite{Kun18,Edv19}.
One of the reasons for the breakdown of the bulk-boundary correspondence in NH systems is that the bulk states are often localized in the vicinity of the boundary (NH skin effect) so that the boundary effects are not described by the bulk Bloch Hamiltonian \cite{Yao18, Kaw18, Rui19, Bor19, Lee19, Hel19, Hof19}. 
Nevertheless, for particular NH symmetry classes, such as the pseudo-Hermitian Hamiltonians, the bulk-boundary correspondence can be established \cite{Kaw19}.

\begin{figure}[t]
\includegraphics[width=1\columnwidth]{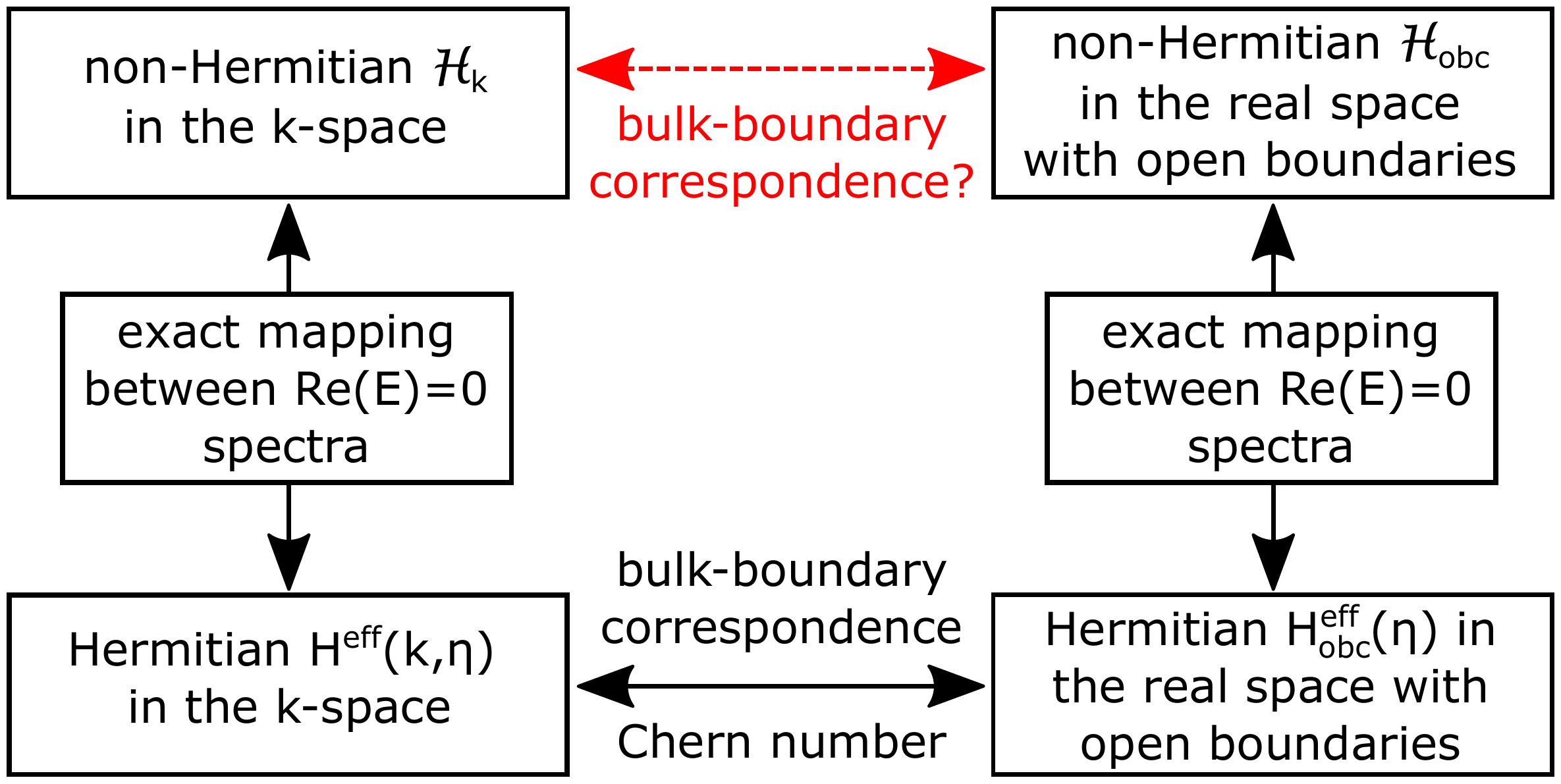}\caption{Schematic representation of the hidden Chern number and bulk-boundary correspondence for the 1D non-Hermitian chiral-symmetric Hamiltonians ${\cal H}_{k}$ satisfying relation (\ref{eq:chir}). The bulk topology of ${\cal H}_{k}$ is described by an effective 2D Hermitian Hamiltonian $H^{\rm eff} (k, \eta)$, where $\eta$ is the imaginary part of the energy. The 2D Hermitian Hamiltonian $H^{\rm eff} (k, \eta)$ supports Chern number as a topological invariant which determines the existence of boundary modes for the corresponding Hermitian Hamiltonian $H^{\rm eff}_{\rm obc}(\eta)$ with open boundary conditions. The $H^{\rm eff}_{\rm obc}(\eta)$  determines the existence of boundary states for the non-Hermitian Hamiltonian ${\cal H}_{\rm obc}$ with open boundary conditions. \label{fig0}}
\end{figure}

From the viewpoint of experiments, particularly relevant 1D NH Hamiltonians can be constructed by considering Hermitian hopping Hamiltonians with onsite gain and loss terms. Assuming chiral-symmetric hopping Hamiltonian this type of NH Hamiltonians ${\cal H}_{k}$ satisfy a special
kind of {\it NH chiral symmetry}
\begin{equation}
{\cal S}{\cal H}_{k}{\cal S}=-{\cal H}_{k}^{\dagger},\label{eq:chir}
\end{equation}
where ${\cal S}$ is a Hermitian unitary operator. Furthermore, in the following we assume that ${\cal S}$ is traceless so that the unit cell contains even number of lattice sites. In the topological classification discussed in Refs.~\cite{Kaw19},
the NH chiral-symmetric Hamiltonians are  discussed and the topological invariant and bulk-boundary correspondence has been established \cite{Zha19,Kaw19}. 
They can also be considered in the framework of pseudo-Hermitian Hamiltonians because $i{\cal H}_{k}$ is a pseudo-Hermitian matrix. In this paper we establish a novel perspective on the topological invariants and bulk-boundary correspondence for these Hamiltonians. Namely, we show that the topology of 1D NH chiral-symmetric Hamiltonian satisfying Eq.~(\ref{eq:chir}) is described by an effective 2D Hermitian Hamiltonian $H^{\rm eff} (k, \eta)$, where $\eta$ is the imaginary part of the energy. Moreover, we show that $H^{\rm eff} (k, \eta)$ supports Chern number as a topological invariant which determines the existence of boundary states also for the non-Hermitian Hamiltonian ${\cal H}_{\rm obc}$ with open boundary conditions via the bulk-boundary correspondence of Hermitian Hamiltonians (see Fig.~\ref{fig0}). Finally, we introduce a minimal model Hamiltonian supporting topologically nontrivial phases in this symmetry class, derive its topological phase diagram and calculate the end states originating from the hidden Chern number. The hidden Chern numbers agree with the topological invariant considered in Ref.~\cite{Kaw19}. 

By inspecting the characteristic polynomial of ${\cal H}_{k}$ satisfying Eq.~(\ref{eq:chir}) we
note that if $\lambda$ is the eigenvalue of ${\cal H}_{k}$ then
also $-\lambda^{\star}$ is. Every square matrix can be easily decomposed
in Hermitian and anti-Hermitian part, namely ${\cal H}_{k}={\cal H}_{k}^{h}+{\cal H}_{k}^{a}$
with ${\cal H}_{k}^{h/a}=\frac{1}{2}\left({\cal H}_{k}\pm{\cal H}_{k}^{\dagger}\right)$.
The chiral relation (\ref{eq:chir}) means that
\begin{eqnarray}
{\cal S}{\cal H}_{k}^{h}{\cal S} & = & -{\cal H}_{k}^{h},\nonumber \\
{\cal S}{\cal H}_{k}^{a}{\cal S} & = & {\cal H}_{k}^{a},
\end{eqnarray}
so ${\cal S}$ anticommutes with Hermitian and commutes with anti-Hermitian
part of the Hamiltonian. Moreover, the NH Hamiltonian ${\cal H}_{k}$ satisfies the
relation with a Hermitian Hamiltonian $H_{k}$ [see Appendix \ref{sec:proof}] 
\begin{equation}
{\cal S}{\cal H}_{k}{\cal S}=-{\cal H}_{k}^{\dagger}\iff{\cal H}_{k}=i{\cal S}H_{k}, \ H^\dag_{k}=H_{k}. \label{eq:equiv}
\end{equation}

For traceless ${\cal S}$ we find that ${\cal H}_{k}$ satisfying
Eq. (\ref{eq:chir}) has a generic block structure in the eigebasis
of ${\cal S}$. Namely 
\begin{equation}
{\cal H}_{k}=\begin{pmatrix}iP_{k} & Q_{k}^{\dagger}\\
Q_{k} & iR_{k}
\end{pmatrix},\label{eq:blocks}
\end{equation}
where $P_{k}$ and $R_{k}$ are $N\times N$ Hermitian matrices and
\begin{equation}
{\cal S}=\begin{pmatrix}\mathbbm{1} & 0\\
0 & -\mathbbm{1}
\end{pmatrix}.\label{eq:Sblocks}
\end{equation}
It is worth noticing that any $2\times2$ real traceless Hamiltonian
${\cal H}_{k}$ can be put in the form of Eq. (\ref{eq:blocks}) so
it satisfies the chiral symmetry (\ref{eq:chir}), as shown in Appendix \ref{sec:Real--case}. 

Now consider the real space version of this Hamiltonian with open boundary conditions ${\cal H}_{{\rm obc}}$. If ${\cal H}_{k}$
satisfies (\ref{eq:chir}) then also ${\cal H}_{{\rm obc}}$ satisfies
it with ${\cal S}_{{\rm obc}}=\mathbbm{1}_{L}\otimes{\cal S}$, where
$L$ is the number of unit cells stacked along the chain. We are interested about the end states of ${\cal H}_{{\rm obc}}$ with zero real part of the energy. Thus, we demand that there exists 
$\eta \in \mathbb{R}$ such that
\begin{equation}
{\cal H}_{{\rm obc}}\left|\psi\right\rangle =i\eta\left|\psi\right\rangle. \label{eq:eig}
\end{equation}
Using (\ref{eq:equiv}) we obtain ${\cal H}_{{\rm obc}}=i{\cal S}_{{\rm obc}}H_{{\rm obc}}$ and $H^\dag_{{\rm obc}}=H_{{\rm obc}}$, so that we get from Eq.~(\ref{eq:eig}) 
\begin{equation}
\left(H_{{\rm obc}}+\eta{\cal S}_{{\rm obc}}\right)\left|\psi\right\rangle =0.\label{eq:eig_eff}
\end{equation}
Notice that this is a  zero-energy eigenproblem for an effective {\it Hermitian} Hamiltonian 
\begin{equation}
H_{{\rm obc}}^{{\rm eff}}\left(\eta\right)\equiv H_{{\rm obc}}+\eta{\cal S}_{{\rm obc}}={\cal S}_{{\rm obc}}\left(\eta-i{\cal H}_{{\rm obc}}\right).\label{eq:heff}
\end{equation}
We can use this to define a $k$-space effective Hermitian Hamiltonian 
\begin{equation}
H^{{\rm eff}}\left(k,\eta\right)\equiv H_{k}+\eta{\cal S}={\cal S}\left(\eta-i{\cal H}_{k}\right).\label{eq:heffk}
\end{equation}
The above Hamiltonian is defined in a two-dimensional $\left(k,\eta\right)$
space and its Chern number is quantized if it is gapped and it can be compactified in $\eta$. A non-trivial
Chern number $C$ of $H^{{\rm eff}}\left(k,\eta\right)$ at half-filling
implies that we have $C$ chiral boundary-modes of $H_{{\rm obc}}^{{\rm eff}}\left(\eta\right)$
crossing the energy gap and it means that we have $C$ solutions of
the eigenproblem (\ref{eq:eig_eff}) or (\ref{eq:eig}). Mapping the non-Hermitian Hamiltonians with open and periodic boundary conditions to the same Hermitian
problem guarantees that if ${\cal H}_{k}$ has gapped real spectrum
then ${\cal H}_{{\rm obc}}$ also has gapped real spectrum meaning
that the non-Hermitian skin effect does not lead to a breakdown of the bulk-boundary correspondence. Moreover, if $H^{{\rm eff}}\left(k,\eta\right)$
is gapped and topological then $H_{{\rm obc}}^{{\rm eff}}\left(\eta\right)$
supports boundary states, and therefore also ${\cal H}_{k}$ is gapped and ${\cal H}_{{\rm obc}}$ supports
end states with zero real part of the energy (see Fig.~\ref{fig0} for the schematic view of this induced bulk-boundary correspondence).

Now what remains is the question of quantization of the Chern number
of $H^{{\rm eff}}\left(k,\eta\right)$. We know that it is quantized
as long as $H^{{\rm eff}}\left(k,\eta\right)$ is periodic in $k$
and $\eta$. Periodicity in $k$ is obvious but in the canonical basis where Eqs.~(\ref{eq:blocks}) and (\ref{eq:Sblocks}) are satisfied we obtain 
\begin{equation}
H^{{\rm eff}}\left(k,-\infty\right)=-H^{{\rm eff}}\left(k,+\infty\right).\label{eq:etainf}
\end{equation}
We can overcome this problem by defining a compactified version of Hamiltonian
$H^{{\rm eff}}\left(k,\eta\right)$ given by
\begin{equation}
H_{{\rm cp}}^{{\rm eff}}\left(k,\eta\right)={\cal R}_{\eta}H^{{\rm eff}}\left(k,\eta\right){\cal R}_{\eta}^{\dagger}.
\end{equation}
where 
\begin{equation}
{\cal R}_{\eta}=\exp\left[i\frac{\pi}{4}\left(1+\tanh\eta\right){\cal G}\right],\quad{\cal G}=\begin{pmatrix}0 & \mathbbm{1}\\
\mathbbm{1} & 0
\end{pmatrix}.
\end{equation} 
This way  ${\cal R}_{-\infty}=\mathbbm{1}$ and ${\cal R}_{+\infty}=i{\cal G}$
so that $H_{{\rm cp}}^{{\rm eff}}\left(k,\eta\right)$ is compactified
in $\eta$ as 
\begin{equation}
H_{{\rm cp}}^{{\rm eff}}\left(k,\eta\to-\infty\right)=H_{{\rm cp}}^{{\rm eff}}\left(k,\eta\to+\infty\right)=-\left|\eta\right|{\cal S}.
\end{equation}
Note that the spectrum of $H_{{\rm cp}}^{{\rm eff}}\left(k,\eta\right)$
and $H^{{\rm eff}}\left(k,\eta\right)$ is the same. Now, the Chern
number $C$ for $H_{{\rm cp}}^{{\rm eff}}\left(k,\eta\right)$ can
be obtained using Kubo formula \cite{Thouless82, Nagaosa10}
\begin{equation}
C=\frac{1}{2 \pi}\int_{-\infty}^{+\infty}d\eta\int_{0}^{2\pi}dk\Omega_{k,\eta},
\end{equation}
where the Berry curvature $\Omega_{k,\eta}$ is given by
\begin{equation}
\Omega_{k,\eta}=\!\!\sum_{{n\le n_{F}\atop m>n_{F}}}\!\!\mathfrak{Im}\frac{2\left\langle \psi_{k,\eta}^{n}\right|\!\partial_{k}H_{{\rm cp}}^{{\rm eff}}\!\left|\psi_{k,\eta}^{m}\right\rangle \!\left\langle \psi_{k,\eta}^{m}\right|\!\partial_{\eta}H_{{\rm cp}}^{{\rm eff}}\!\left|\psi_{k,\eta}^{n}\right\rangle }{\left(E_{k,\eta}^{(n)}-E_{k,\eta}^{(m)}\right)^{2}}.
\end{equation}
Here $\left|\psi_{k,\eta}^{p}\right\rangle $ are eigenstates of $H_{{\rm cmp}}^{{\rm eff}}\left(k,\eta\right)$ (sorted in ascending order of eigenenergy) and $n_{F}$ is the number of occupied bands.

\begin{figure}
\includegraphics[width=1\columnwidth]{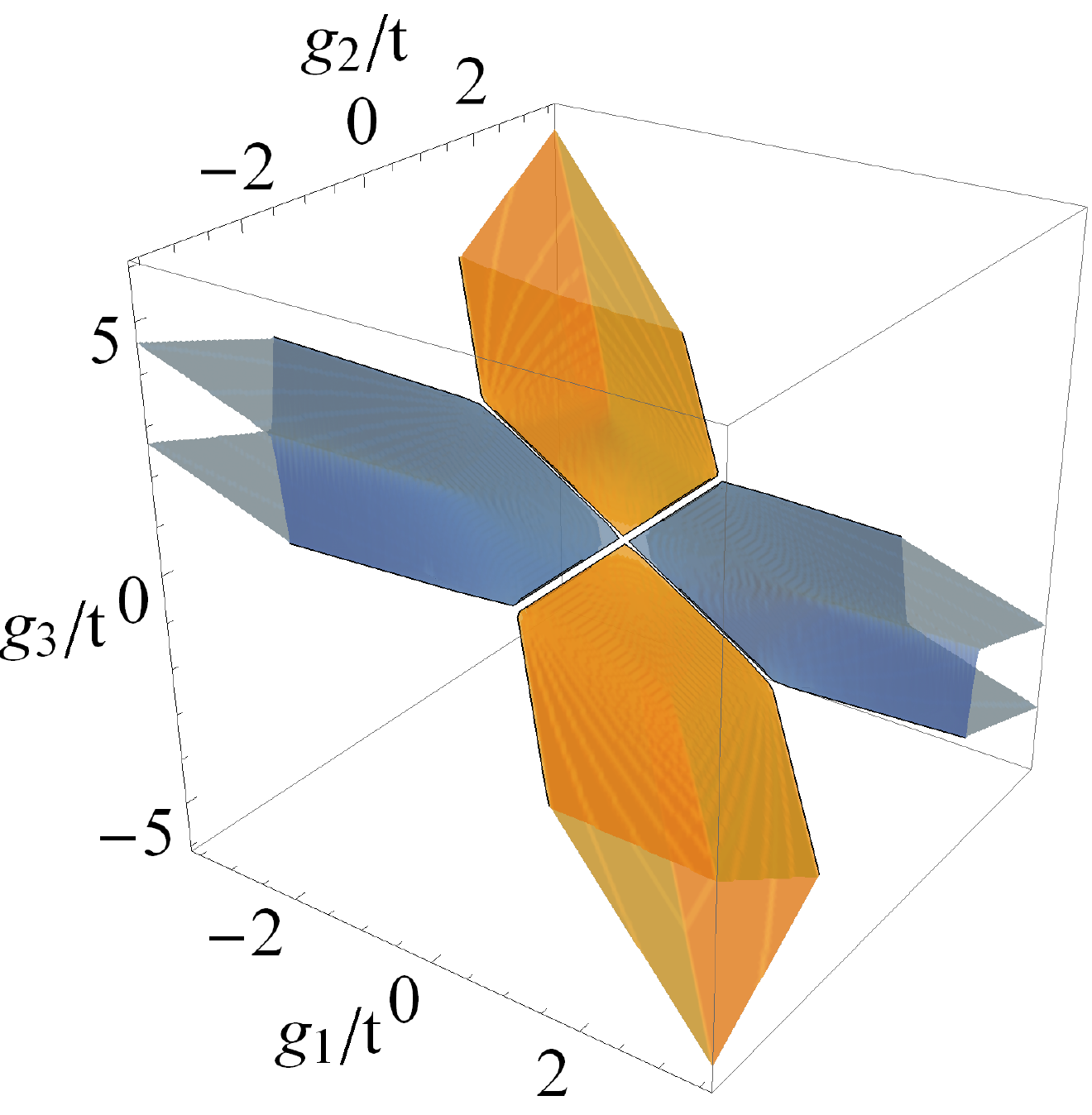}\caption{Phase diagram of the effective Hermitian model (\ref{eq:Hameff}).
Inside colored surfaces there are gapped phases with Chern numbers
$C=-1$ (orange) and $C=0$ (blue). Outside colored surfaces there
are gapless phases with indirect gap closings. \label{fig1}}
\end{figure}

\begin{figure}
\includegraphics[width=1\columnwidth]{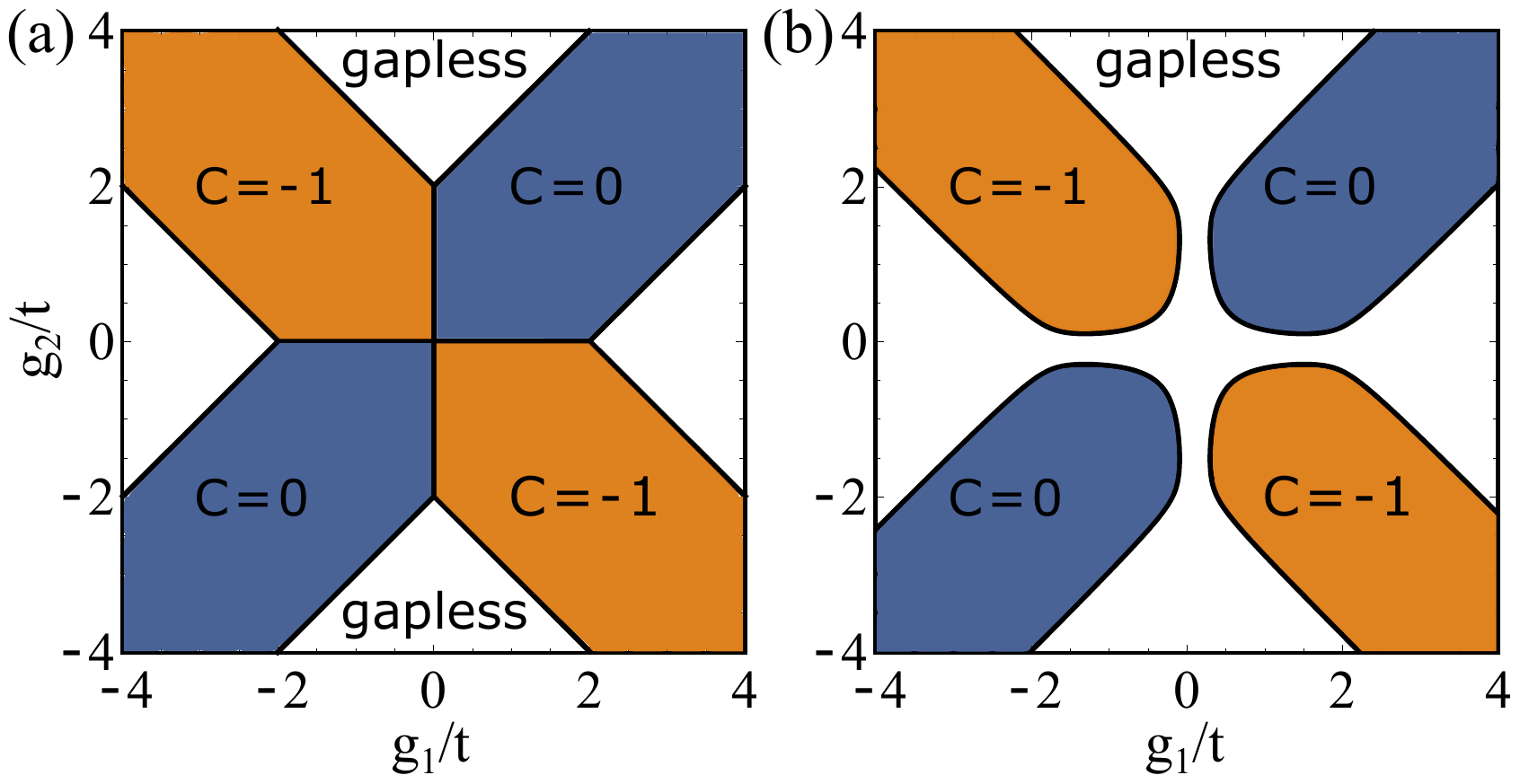}\caption{Phase diagram of the effective Hermitian model (\ref{eq:Hameff})
for (a) $g_{3}=-g_{1}$ and (b) $g_{3}=-g_{1}+0.2t$. Inside colored
surfaces there are gapped phases with Chern numbers $C=-1$ (orange)
and $C=0$ (blue). Outside colored regions there are gapless phases
with indirect gap closings. \label{fig2}}
\end{figure}

In what follows we focus on a minimal tight-binding model with uniform hopping $t$ supporting nontrivial phases. In this model we have a four-site unit cell with gain and loss terms  $g_{1}$, $g_{2}$, $g_{3}$ and $g_{4}$, so that the NH Hamiltonian is
\begin{equation}
{\cal H}_{k}=\begin{pmatrix}ig_{1} & t & 0 & te^{-ik}\\
t & ig_{2} & t & 0\\
0 & t & ig_{3} & t\\
te^{ik} & 0 & t & ig_{4}
\end{pmatrix}.\label{eq:Ham}
\end{equation}
Without loss of generality we can assume that 
\begin{equation}
g_{4}=-g_{1}-g_{2}-g_{3}.
\end{equation}
This model satisfies chiral relation (\ref{eq:chir})
with ${\cal S}=\mathbbm{1}\otimes\sigma_{z}$. From Eq. (\ref{eq:heffk})
we find that the effective Hermitian Hamiltonian for our model is
\begin{equation}
H^{{\rm eff}}\left(k,\eta\right)=\begin{pmatrix}\eta+g_{1} & -it & 0 & -ite^{-ik}\\
it & -\eta-g_{2} & it & 0\\
0 & -it & \eta+g_{3} & -it\\
ite^{ik} & 0 & it & -\eta-g_{4}
\end{pmatrix},\label{eq:Hameff}
\end{equation}
and similarly from Eq. (\ref{eq:heff}) we can get $H_{{\rm obc}}^{{\rm eff}}\left(\eta\right)$.
The phase diagram of $H^{{\rm eff}}\left(k,\eta\right)$ is given
in Fig.~\ref{fig1}. The gapped phases are located mainly in the plane
$g_{3}=-g_{1}$ and we find two non-trivial ones with hidden Chern
numbers $C=-1$ along direction $g_{1}=-g_{2}$ and two trivial ones
with $C=0$ along direction $g_{1}=g_{2}$. For better clarity Fig.~\ref{fig2} shows two-dimensionl phase diagrams in the planes $g_{3}=-g_{1}$
and $g_{3}=-g_{1}+0.2t$, see also Appendix \ref{sec:Phase-diagrams}.

\begin{figure}
\includegraphics[width=1\columnwidth]{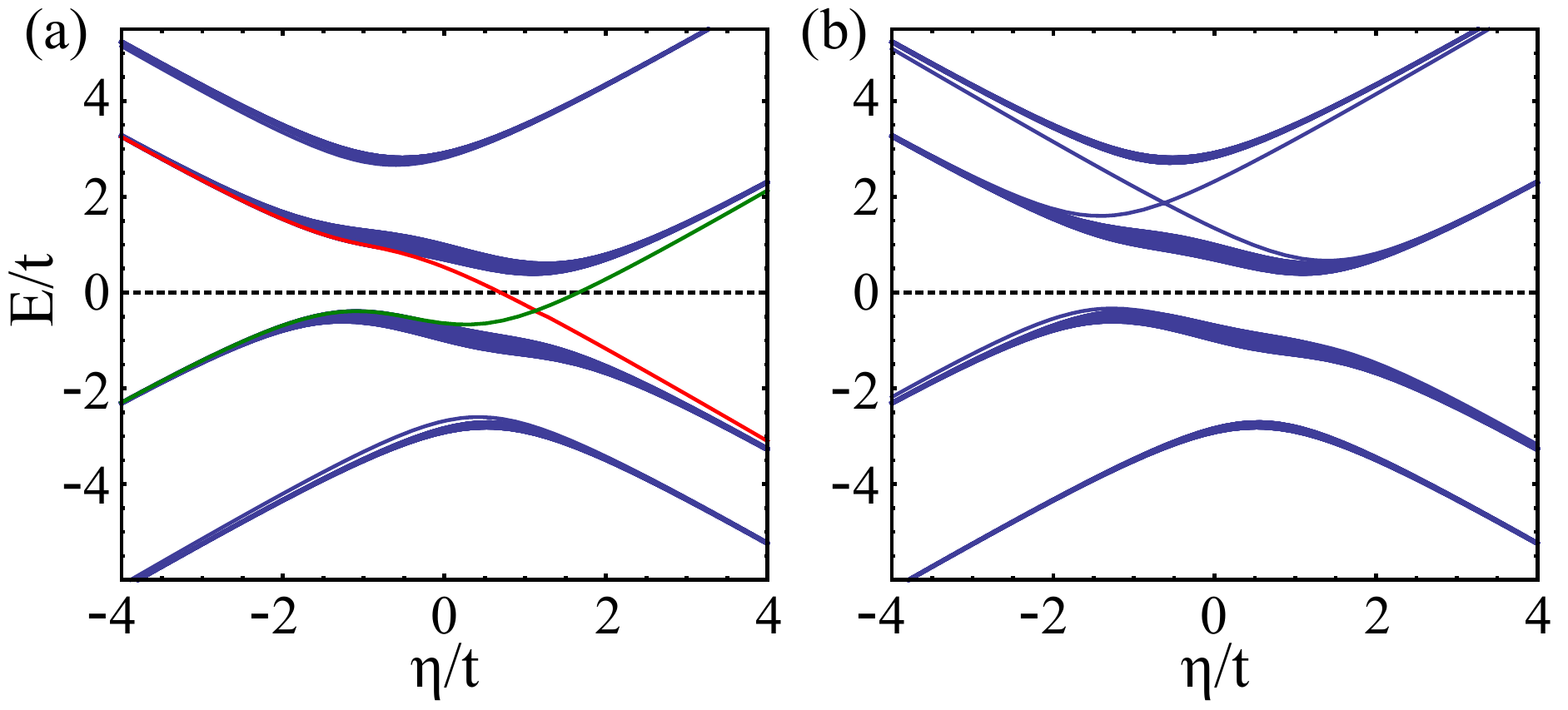}\caption{Spectrum of the Hermitian Hamiltonian $H_{{\rm obc}}^{{\rm eff}}\left(\eta\right)$ as function of $\eta$ for (a) $g_{1}=2t$, $g_{2}=-t$ ($C=-1$) and (b)
$g_{1}=-2t$, $g_{2}=-t$ ($C=0$). In both cases $g_{3}=-g_{1}$. Zero energy level is marked with
dashed line and the boundary states localized at the opposite ends of the chain are shown in red and green. \label{fig3}}
\end{figure}

\begin{figure}
\includegraphics[width=1\columnwidth]{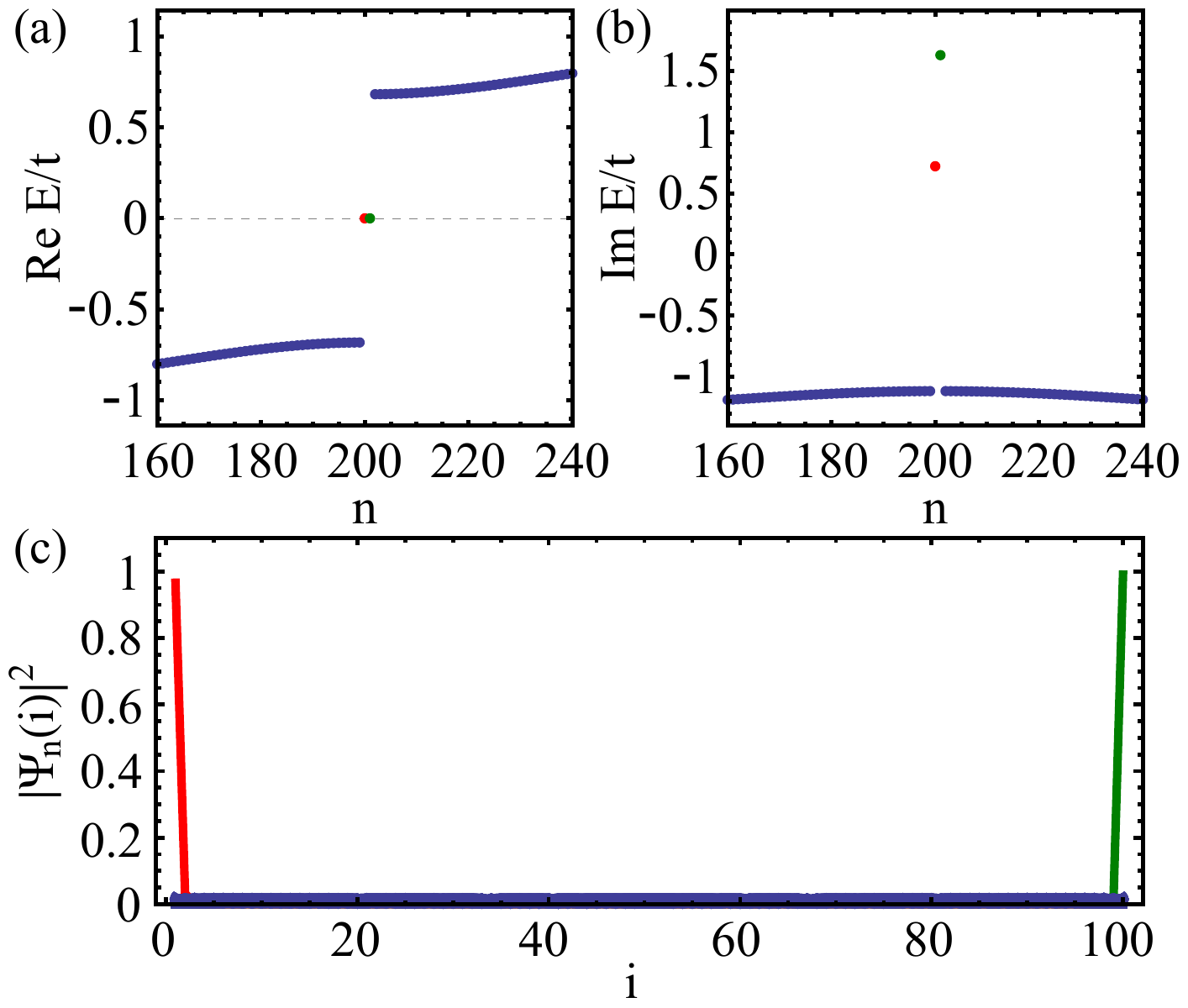}\caption{(a-b) Real and imaginary parts of the eigenenergy spectrum of the non-Hermitian Hamiltonian
${\cal H}_{{\rm obc}}$ (\ref{eq:Ham}) with open boundary condition. States are ordered by increasing real part of the energy and $n$ enumerates states. (c) Local density of states for
each eigenstate.  Parameters are $g_{1}=t$, $g_{2}=-2.2t$ and $g_{3}=-g_{1}+0.2t$, and the
system length is $L=100$ sites. Red/green dots or lines correspond to the
boundary states localized at the opposite ends of the chain. \label{fig4}}
\end{figure}

In Fig.~\ref{fig3} we show the spectra of the Hamiltonian $H_{{\rm obc}}^{{\rm eff}}\left(\eta\right)$ with open boundary conditions as a function of $\eta$ for the gapped phases  shown in Fig.~\ref{fig2}(a). In the non-trivial phase {[}Fig.~\ref{fig3}(a){]}
we can see that the gap around zero energy is crossed by two states
connecting valence and conduction band. Each of them cross the zero
energy level once and values of $\eta$ at which this happen correspond
to two solutions of Eq.~(\ref{eq:eig}), so that in the case of the original non-Hermitian Hamiltonian ${\cal H}_{{\rm obc}}$ the real part of the eigenenergy is zero. In the
trivial case {[}Fig. \ref{fig3}(b){]} these states are missing. 

Fig.~\ref{fig4} shows energy spectra for a topologically nontrivial non-Hermitian system with open boundary conditions. We find two end states with zero real part of the energy and the imaginary parts of the energies stick out of the bulk spectrum.
As one can see from Fig. \ref{fig4}(c) the states with zero real part of the energy are strongly localized
at the left or right end of the chain. The rest of the states are delocalized in the bulk.

We have also computed the topological invariant for the model Hamiltonian [Eq.~(\ref{eq:Ham})] using the approach discussed in  Ref.~\cite{Kaw19}. We find that nontrivial invariants appear exactly in the same gapped regions where our hidden Chern number is non-zero. Therefore, we conclude that the hidden Chern number gives an alternative perspective on topological invariants and bulk-boundary correspondence of NH chiral-symmetric Hamiltonians.

In summary, we have shown that one-dimensional non-Hermitian chiral-symmetric
models support a hidden Chern number as a topological invariant.  The Chern number determines the number of end states with zero real part of the energy and the end states are immune
to non-Hermitian skin effect. Moreover, we have calculated the topological phase diagram and end states for a minimal $4\times4$ gain and loss model that supports a nontrivial topological phase. Our approach gives a new perspective on the topological invariants and bulk-boundary correspondence of non-Hermitian systems and the idea can be generalized to various dimensions and symmetries. Although there exists several proposals for realizing topological invariants of Hermitian systems in lower dimensional non-Hermitian systems \cite{Zen19, Vishwanath19}, the idea to utilize the imaginary part of the energy as a new dimension has not been explored so far.  We point out that also gapless phases can support hidden Chern numbers, and in particular the gapless phase of the non-Hermitian Hamiltonian (\ref{eq:Ham}) carries hidden Chern number $C=-1$ in the vicinity of topologically nontrivial gapped phase -- leading to localized topological end states, see Appendix \ref{sec:Gapless-phases}. The end states can be utilized for example in laser modes which have topologically protected frequency stability.

\begin{acknowledgments}
The work is supported by the Foundation for Polish Science through
the IRA Programme co-financed by EU within SG OP Programme.
\end{acknowledgments}

\appendix

\section{Proof of equivalence (3)\label{sec:proof}}

Assume that there exist a Hermitian unitary operator ${\cal S}$ such
that
\begin{equation}
{\cal S}{\cal H}_{k}{\cal S}=-{\cal H}_{k}^{\dagger}. \label{eq:chir2}
\end{equation}
Then we can write
\begin{equation}
{\cal H}_{k}=i{\cal S}H_{k}
\end{equation}
where
\begin{equation}
H_{k}=-i{\cal S}{\cal H}_{k}.
\end{equation}
Now $H_{k}$ is Hermitian because
\begin{equation}
H_{k}^{\dagger}=i{\cal H}_{k}^{\dagger}{\cal S}=-i{\cal S}{\cal H}_{k}=H_{k}.
\end{equation}

On the other hand, if we assume that 
\begin{equation}
{\cal H}_{k}=i{\cal S}H_{k},
\end{equation}
and ${\cal S}$ is Hermitian unitary and $H_{k}$ is Hermitian, then ${\cal S}{\cal H}_{k}{\cal S}=-{\cal H}_{k}^{\dagger}$.

\begin{figure}[t!]
\includegraphics[width=1 \columnwidth]{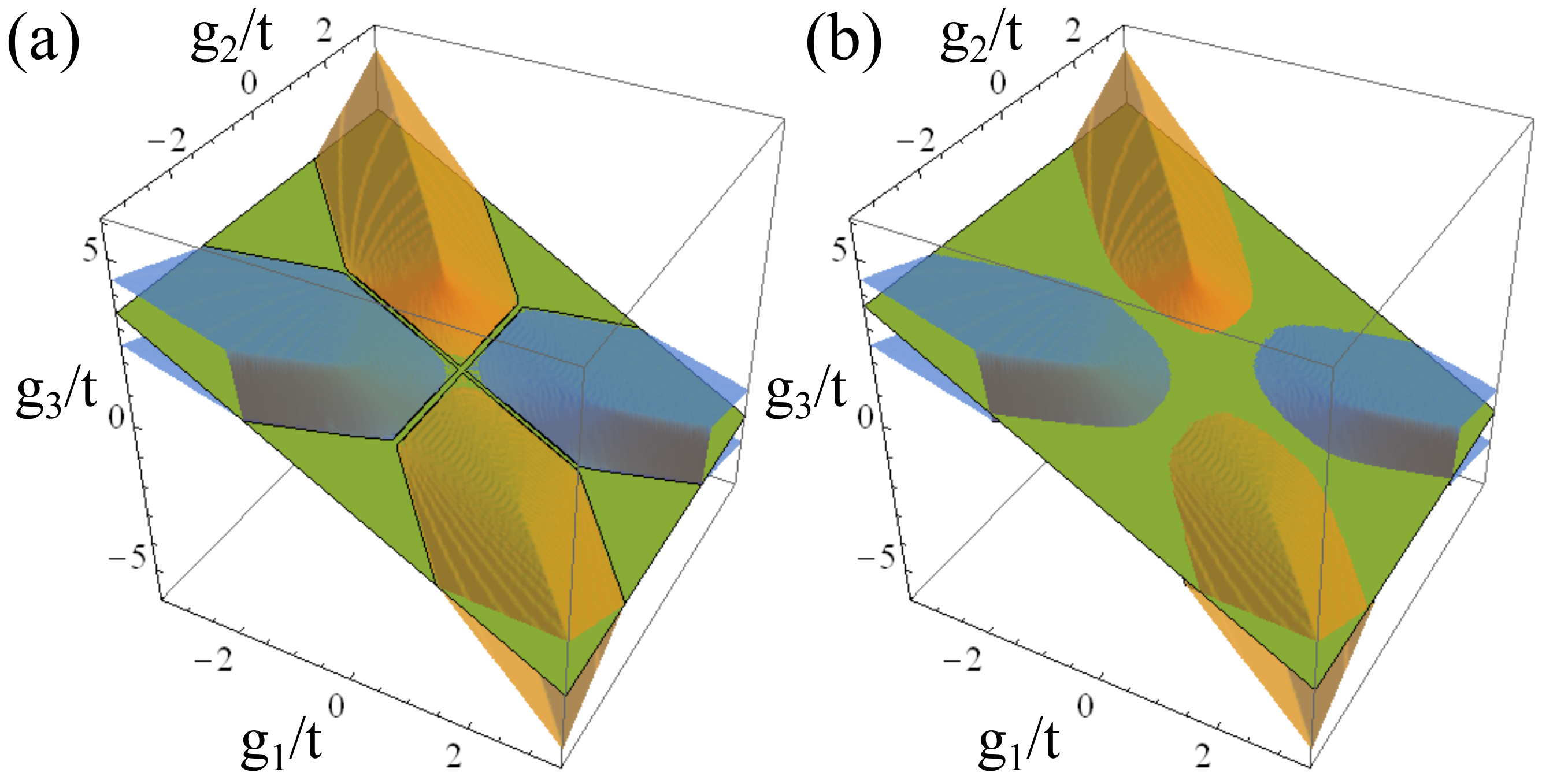}
\caption{Phase diagram of the effective Hermitian model [Eq. (18) of the main text].
Inside colored surfaces there are gapped phases with Chern numbers
$C=-1$ (orange) and $C=0$ (blue). Outside colored surfaces there
are gapless phases with indirect gap closings. Green planes indicate surfaces
parametrized by  (a) $g_{3}=-g_{1}$ and (b) $g_{3}=-g_{1}+0.2t$. 
\label{fig7}}
\end{figure}

\section{Real $2\times2$ case\label{sec:Real--case}}

A real traceless $2\times2$ Hamiltonian has a form of 
\begin{equation}
{\cal H}_{k}=\begin{pmatrix}a_{k} & b_{k}\\
c_{k} & -a_{k}
\end{pmatrix},
\end{equation}
where $a_{k}$, $b_{k}$ and $c_{k}$ are real functions. By a unitary
transformation ${\cal U}=\exp\left[i\frac{\pi}{3\sqrt{3}}(\sigma_{x}+\sigma_{y}+\sigma_{z})\right]$,
we get
\begin{equation}
{\cal H}'_{k}={\cal U}^{\dagger}{\cal H}_{k}{\cal U}=\begin{pmatrix}\frac{i}{2}\left(b_{k}-c_{k}\right) & a_{k}-\frac{i}{2}\left(b_{k}+c_{k}\right)\\
a_{k}+\frac{i}{2}\left(b_{k}+c_{k}\right) & -\frac{i}{2}\left(b_{k}-c_{k}\right)
\end{pmatrix}.
\end{equation}
The block structure of ${\cal H}'_{k}$ is obviously the one of Eq.
(4) of the main text and we find that $\sigma_{z}{\cal H}'_{k}\sigma_{z}=-\left({\cal H}'_{k}\right)^{\dagger}$.

\section{Phase diagrams\label{sec:Phase-diagrams}}

To establish a clear link between phase diagrams shown in Figs. 2 and 3 in the main text we present
in Fig. \ref{fig7} three dimensional phase diagrams cut by the planes of $g_{3}=-g_{1}$ and  $g_{3}=-g_{1}+0.2t$.
Curves along which they cut the branches of the phase diagram are the boundaries of the phases shown in Fig. 3 of
the main text. 

\begin{figure}[b!]
\includegraphics[width=1\columnwidth]{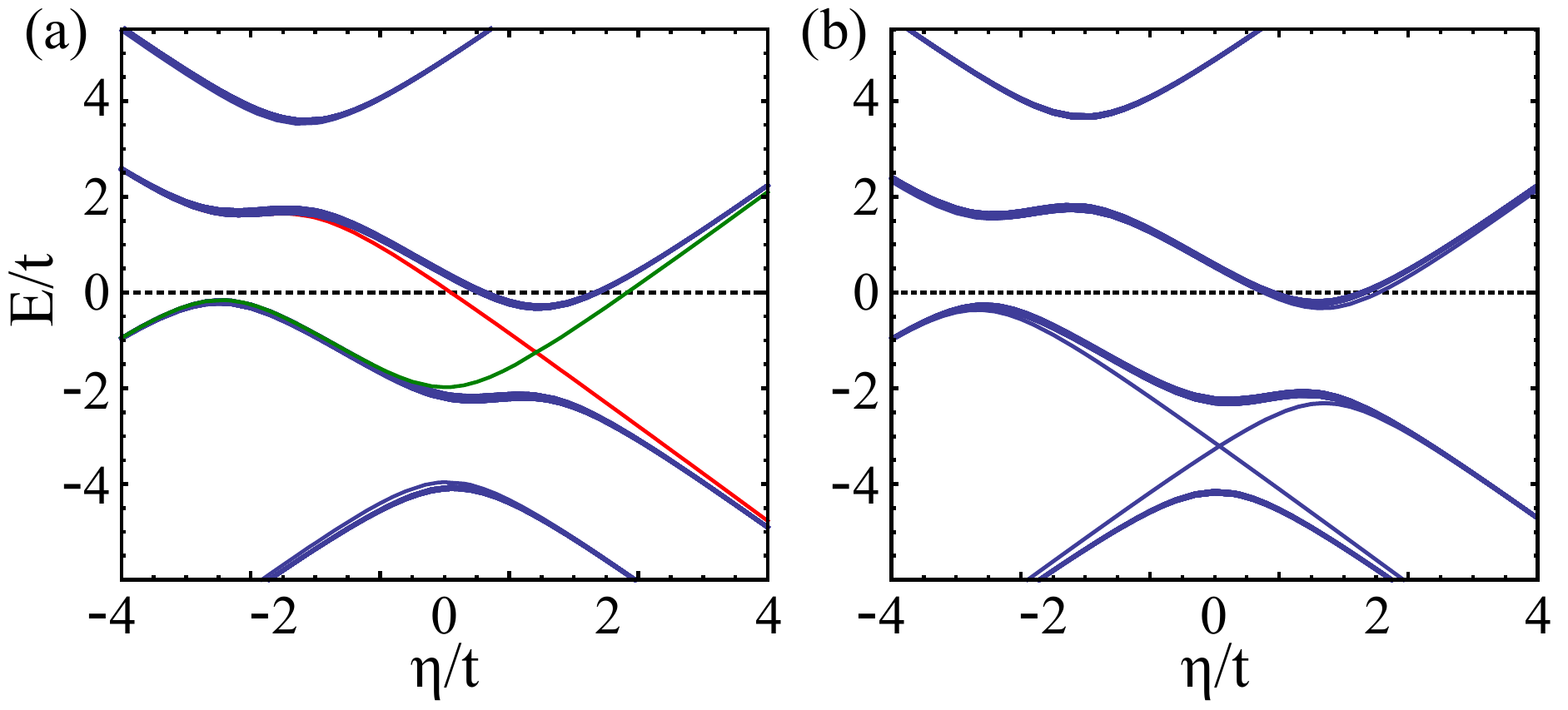}
\caption{(a) Spectra of the Hermitian Hamiltonian $H_{{\rm obc}}^{{\rm eff}}\left(\eta\right)$ in
a nontrivial gapless phase as function of $\eta$ for $g_{1}=4t$, $g_{2}=-1.8t$ and $g_{3}=-g_{1}+0.5t$ ($C=-1$). (b) Same for trivial phase with $g_{1}=4t$, $g_{2}=1.5t$ and $g_{3}=-g_{1}+0.5t$ ($C=0$).
Zero energy level is marked with dashed line and boundary states are shown in red and green. There is a continuum of bulk states crossing the zero energy because the real part of ${\cal H}{}_{k}$ is gapless.\label{fig5}}
\end{figure}

\section{Gapless phases\label{sec:Gapless-phases}}

\begin{figure}
\includegraphics[width=1\columnwidth]{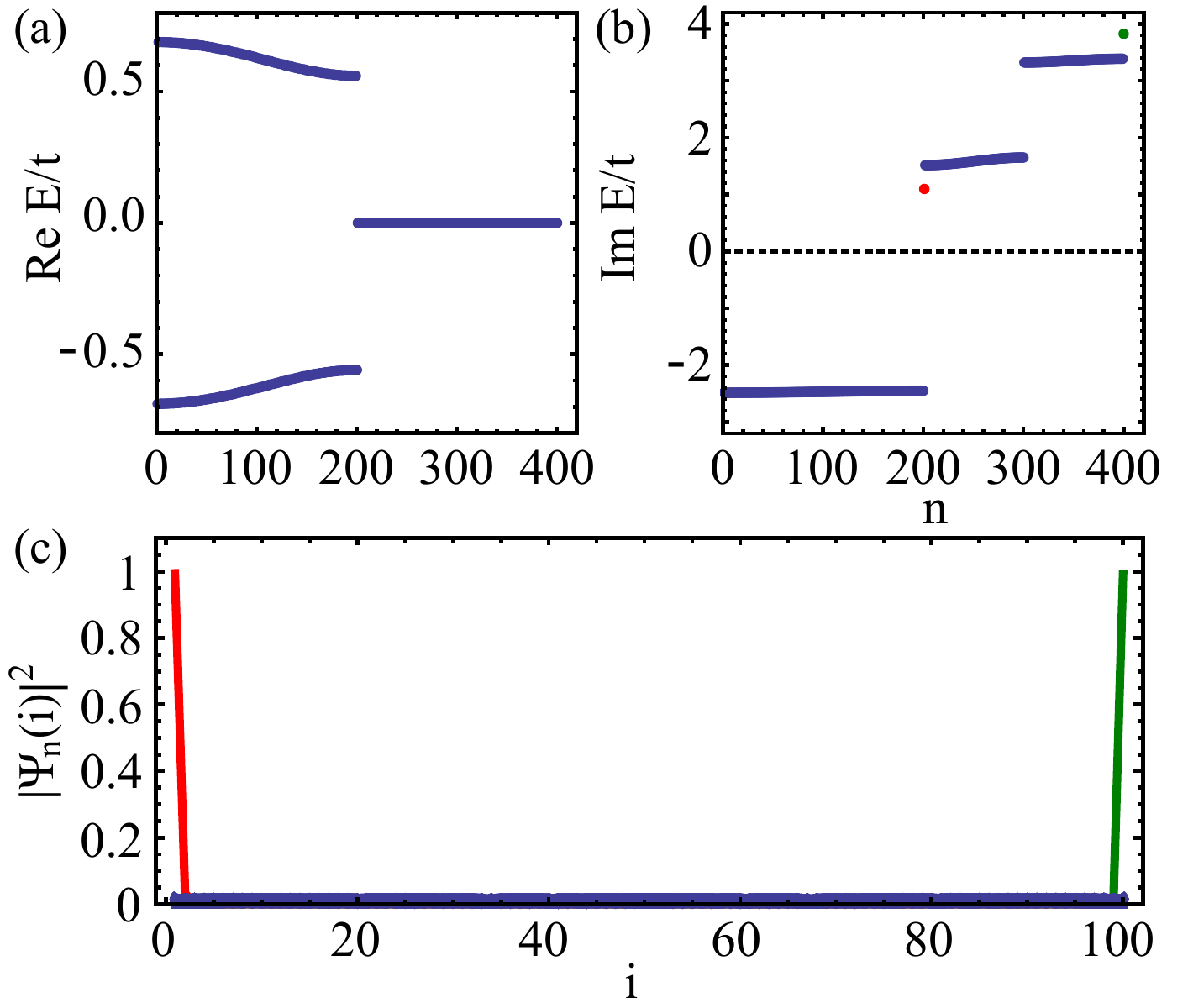}
\caption{(a-b) Real and imaginary parts of the eigenenergy spectrum of the non-Hermitian Hamiltonian
${\cal H}_{{\rm obc}}$ with open boundary condition. States are ordered by increasing imaginary part of the energy and $n$ enumerates states. (c) Local density of states for
each eigenstate. Parameters are $g_{1}=4t$, $g_{2}=-1.8t$ and $g_{3}=-g_{1}+0.5t$, and the
system length is $L=100$ sites. Red/green dots or lines correspond to the
boundary states localized at the opposite ends of the chain.
 \label{fig6}}
\end{figure}

Gapless phases shown in Fig. 2 of the main text also carry
hidden Chern numbers because the Chern number can change only in a band touching point. In a gapless phase close to $C=-1$ (or $C=0$) gapped phase we can still define
a Chern number which takes the same value. The corresponding spectra for the Hermitian and non-Hermitian Hamiltonians with open boundary conditions are shown in Figs.~\ref{fig5} and \ref{fig6}. There is a continuum
of bulk states crossing zero-energy level which is consistent with
real part of ${\cal H}{}_{k}$ being gapless. Moreover,  in the non-trivial case of  we additionally
see boundary states [see Fig.~\ref{fig5}(a) and \ref{fig6}]. The end states are localized and the imaginary parts of the energies are in the gaps of the bulk spectrum [see \ref{fig6}(b)].

\end{document}